\documentclass[preprint]{aastex61}

\usepackage{float}

\def\Gthontw {G031.24$-$00.11}
\def\Gthtwse {G032.79+00.19}
\def\Gfozefo {G040.42+00.70}
\def\Gfozesi {G040.62$-$00.13}
\def\Gfotwze {G042.03+00.19}
\def\Gfothon {G043.16+00.01}
\def\Gfoeisi {G048.60+00.02}
\def\Gfonitw {G049.26+00.31}
\def\Gfonifo {G049.41+00.32}

\def\Gsezeon {G070.18$+$01.74}

\def\hho   {H$_2$O}
\def\meth  {CH$_3$OH}

\def\uas       {$\mu$as}
\def\deg       {$^\circ$}
\def\kms       {km~s$^{-1}$}
\def\masy      {mas~yr$^{-1}$}

\def\Msunyr    {\ifmmode {M_\odot~\rm yr^{-1}}\else {$M_\odot~\rm yr^{-1}$}\fi}

\def\decdeg    {\ifmmode{{\rlap.}^{\circ}} \else ${\rlap.}^{\circ}$\fi}

\def\VLSR  {$V_{\rm LSR}$}

\def\hho   {H$_2$O}
\def\meth  {CH$_3$OH}
\def\AIPS  {$\mathcal{AIPS}$}

\def\mux   {\ifmmode {\mu_x}\else {$\mu_x$}\fi}
\def\muy   {\ifmmode {\mu_y}\else {$\mu_y$}\fi}
\def\mura  {\ifmmode {\mu_{\alpha}}\else {$\mu_{\alpha}$}\fi}
\def\mude  {\ifmmode {\mu_{\delta}}\else {$\mu_{\delta}$}\fi}

\def\dx    {$\Delta x$}
\def\dy    {$\Delta y$}

\def\Vo    {\ifmmode {V^{Std}_\odot}\else {$V^{Std}_\odot$}\fi}
\def\Uo    {\ifmmode {U^{Std}_\odot}\else {$U^{Std}_\odot$}\fi}
\def\Wo    {\ifmmode {W^{Std}_\odot}\else {$W^{Std}_\odot$}\fi}
\def\VH    {\ifmmode {V^H_\odot}\else {$V^H_\odot$}\fi}
\def\UH    {\ifmmode {U^H_\odot}\else {$U^H_\odot$}\fi}
\def\WH    {\ifmmode {W^H_\odot}\else {$W^H_\odot$}\fi}
\def\V     {\ifmmode {V_\odot}\else {$V_\odot$}\fi}
\def\U     {\ifmmode {U_\odot}\else {$U_\odot$}\fi}
\def\W     {\ifmmode {W_\odot}\else {$W_\odot$}\fi}
\def\VGC   {\ifmmode {V_\odot^{GC}}\else {$V_\odot^{GC}$}\fi}
\def\UGC   {\ifmmode {U_\odot^{GC}}\else {$U_\odot^{GC}$}\fi}
\def\WGC   {\ifmmode {W_\odot^{GC}}\else {$W_\odot^{GC}$}\fi}

\def\Vs    {\ifmmode {V_s}\else {$V_s$}\fi}
\def\Us    {\ifmmode {U_s}\else {$U_s$}\fi}
\def\Ws    {\ifmmode {W_s}\else {$W_s$}\fi}

\def\Vsbar {\ifmmode {\overline{V_s}}\else {$\overline{V_s}$}\fi}
\def\Usbar {\ifmmode {\overline{U_s}}\else {$\overline{U_s}$}\fi}
\def\Wsbar {\ifmmode {\overline{W_s}}\else {$\overline{W_s}$}\fi}

\newbox\grsign \setbox\grsign=\hbox{$>$} \newdimen\grdimen \grdimen=\ht\grsign
\newbox\laxbox \newbox\gaxbox
\setbox\gaxbox=\hbox{\raise.5ex\hbox{$>$}\llap
     {\lower.5ex\hbox{$\sim$}}}\ht1=\grdimen\dp1=0pt
\setbox\laxbox=\hbox{\raise.5ex\hbox{$<$}\llap
     {\lower.5ex\hbox{$\sim$}}}\ht2=\grdimen\dp2=0pt

\def\lax{\mathrel{\copy\laxbox}}
\shorttitle{Inner perseus arm}
\shortauthors{Zhang et al.}
\begin{document}

\title{
Parallaxes for star forming regions in the inner Perseus spiral arm
}

\correspondingauthor{Bo Zhang}
\email{zb@shao.ac.cn}

\author[0000-0003-1353-9040]{Bo Zhang}
\affiliation{Shanghai Astronomical Observatory, Chinese Academy of Sciences\\
80 Nandan Road, Shanghai 200030, China}

\author{Mark J.  Reid}
\affiliation{Center for Astrophysics~$\vert$~Harvard \& Smithsonian\\
60 Garden Street, Cambridge, MA 02138, USA}

\author{Lian Zhang}
\affiliation{Shanghai Astronomical Observatory, Chinese Academy of Sciences\\
80 Nandan Road, Shanghai 200030, China}

\author{Yuanwei Wu}
\affiliation{
National Time Service Center, Chinese Academy of Sciences\\
Xi'an 710600, China
}
\affil{
Mizusawa VLBI Observatory, National Astronomical Observatory of Japan\\
 Mitaka, Tokyo 181-8588, Japan
}

\author{Bo Hu}
\affiliation{
School of Astronomy and Space Science, Nanjing University\\
22 Hankou Road, Nanjing  210093, China}

\author{ Nobuyuki Sakai}
\affil{
Mizusawa VLBI Observatory, National Astronomical Observatory of Japan\\
 Mitaka, Tokyo 181-8588, Japan
}

\author{Karl M. Menten}
\affiliation{Max-Plank-Institut f\"ur Radioastronomie\\
Auf dem H\"ugel 69, 53121 Bonn, Germany}

\author{Xingwu Zheng}
\affiliation{
School of Astronomy and Space Science, Nanjing University\\
22 Hankou Road, Nanjing  210093, China}

\author{Andreas Brunthaler}
\affiliation{Max-Plank-Institut f\"ur Radioastronomie\\
Auf dem H\"ugel 69, 53121 Bonn, Germany}

\author{Thomas M.  Dame}
\affiliation{Harvard-Smithsonian Center for Astrophysics\\
60 Garden Street, Cambridge, MA 02138, USA}

\author{Ye Xu}
\affiliation{
Purple Mountain Observatory, Chinese Academy of Sciences\\
Nanjing 210008, China}



\begin{abstract}

We report trigonometric parallax and proper motion measurements of 6.7-GHz \meth\ 
and 22-GHz \hho\ masers in eight high-mass star-forming regions (HMSFRs) 
based on VLBA observations as part of the BeSSeL Survey.  The distances of these 
HMSFRs combined with their Galactic coordinates, radial velocities, and proper motions,
allow us to assign them to a segment of the Perseus arm with $\ell$~ $\lax$ ~70\deg.
These HMSFRs are clustered in Galactic longitude
from $\approx30$\deg\ to $\approx50$\deg\, neighboring a dirth of such sources
between longitudes $\approx50$\deg\ to $\approx90$\deg.

\end{abstract}


\keywords{astrometry -- Galaxy: fundamental parameters -- Galaxy:
Kinematics and dynamics -- masers -- stars: formation -- techniques:
high angular resolution}



\section{INTRODUCTION}
\label{sec:intro}

Since the Perseus arm has been proposed as one of two major spiral arms
of the Milky Way
\citep{2000A&A...358L..13D,2005ApJ...630L.149B,2009PASP..121..213C}, it
is especially important to study its structure and kinematics.  {\it
Gaia} --- the successor of the {\it Hipparcos} optical astrometry
satellite --- is expected to revolutionize our understanding of the
structure and kinematics of the Milky Way.   However, strong
interstellar extinction, in particular from spiral arms in the inner
Galaxy, will severely limit {\it Gaia} and any optical observations of
spiral structure.  Additionally, systematic offsets might exist between
stars and gas associated with spiral arms
\citep{1969ApJ...158..123R,1972A&A....17..468M}.  Very Long Baseline
Interferometry (VLBI) at radio wavelengths, with angular resolution
better than a milli-arcsecond, can provide astrometric accuracy of
$\approx$ 10~\uas, which is comparable or better than the goal of {\it
Gaia}~\citep{2014ARA&A..52..339R}.  Over the last decade, VLBI parallax
measurements for masers in high-mass star-forming regions (HMSFRs) have
been tracing spiral arms in the first three quadrants of the Milky
Way~\citep{2013ASPC..476...81H,2014ApJ...783..130R} and have
demonstrated the capability to measure accurate distances up to 20
kpc~\citep{2017Sci...358..227S}.

The key to a better understanding of spiral arms is to increase the
number of reliable arm tracers (e.g., HMSFRs) with accurate distances.
We are using the NRAO\footnote{The National Radio Astronomy Observatory
is a facility of the National Science Foundation operated under
cooperative agreement by Associated Universities, Inc} Very Long
Baseline Array (VLBA) to carry out a key science project, the Bar and
Spiral Structure Legacy (BeSSeL)
Survey\footnote{\url{http://bessel.vlbi-astrometry.org/}}, to measure
trigonometric parallaxes and proper motions for hundreds of 22 GHz \hho\
and 6.7/12.2 GHz \meth\ maser sources associated with HMSFRs.  The
Perseus arm at $\ell$~$\lax$~70\deg (hereafter the inner Perseus arm) is
located far from the Sun and only a small number of accurate parallax
distances have been determined~\citep{2013ApJ...775...79Z}.  In this
paper, we report trigonometric parallax measurements for two 22 GHz
\hho\ masers and six 6.7 GHz \meth\ masers in the inner Perseus arm.
Sources in the outer Perseus arm will be reported in Sakai et al.
(2019, in preparation).

\section{OBSERVATIONS AND CALIBRATION PROCEDURES}
\label{sec:obs}

The VLBA program names and epochs of our observations are shown in
Table~\ref{tab:obs}.  Table~\ref{tab:src} lists the observed source
positions, intensities, source separations, reference maser radial
velocities in the Local Standard of Rest frame (\VLSR), and
interferometer restoring beams.  For all these sources, the amplitude of
the parallax signature in Declination was considerably smaller than for
Right Ascension.  Therefore, for the 6.7 GHz \meth\ masers, four
observing epochs were selected to optimally sample the peaks of the
sinusoidal parallax signature in Right Ascension over one year,
maximizing the sensitivity of parallax detection and ensuring that the
parallax and proper motion signatures are uncorrelated.  For 22 GHz
\hho\ masers, six epochs were observed over one year to allow a parallax
measurement with less than 1 year of data, because \hho\ masers spots
can have shorter lifetimes.  The first epoch for BR198L suffered serious
degradation owing to faulty digital baseband converter firmware and was
not used.

Our general observing setup and calibration procedures are described in
\citet{2009ApJ...693..397R}; here we discuss only aspects of the
observations that are specific to the maser sources presented in this
paper. At each epoch, the observations consisted of four 0.5-hour
``geodetic blocks'' used to calibrate and remove unmodeled atmospheric
propagation delays and determine clock parameters.  For the 6.7-GHz
observations, dual-frequency ``geodetic block'' observations were used,
in order to allow separation of dispersive ionospheric delays from
non-dispersive delays.   See the ``supplementary text'' in
\citet{2016SciA....2E0878X} for a details.

Three $\approx$ 1.5-hour periods of phase-referenced observations were
inserted between the blocks.  In these observations, we cycled between
the target maser and several background sources, switching sources every
$\approx$ 30 or 60 seconds for \hho\ or \meth\ masers, respectively.
The typical on-source integration time per epoch for a maser source and
each background source were typically 0.8 hour and 0.2 hour,
respectively.

In the phase-referenced observations, we used four adjacent intermediate
frequency (IF) bands of 16 MHz, each in both right and left circular
polarization (RCP and LCP); the second band contained the maser signals.
The data correlation was performed with the DiFX\footnote{DiFX: A
software Correlator for VLBI using Multiprocessor Computing
Environments, is developed as part of the Australian Major National
Research Facilities Programme by the Swinburne University of Technology
and operated under licence} software correlator
\citep{2007PASP..119..318D} in Socorro, NM, with 1000 and 2000 spectral
channels for BR149  and BR198, respectively, yielding channel spacings
of 0.36 and 0.11 \kms\ for 6.7 GHz \meth\ and 22 GHz \hho\ masers,
respectively.  We observed three International Celestial Reference Frame
(ICRF) sources~\citep{1998AJ....116..516M}, near the beginning, middle
and end of the phase-referencing observations in order to monitor delay
and electronic phase differences among the observing bands.  

The data reduction was conducted using the NRAO's Astronomical Image
Processing System (\AIPS) together with scripts written in ParselTongue
\citep{2006ASPC..351..497K}.  Since in our case the masers are much
stronger than the background sources, we used a spectral channel with
strong and relatively compact maser emission as the interferometer phase
reference.  After separate calibration of the polarized bands, we
combined the RCP and LCP bands to form Stokes I and imaged the continuum
emission of the background sources using the \AIPS\ tasks {\it IMAGR}.
For the masers, we also formed Stokes I and then imaged the emission in
each spectral channel.   Finally, we fitted elliptical Gaussian
brightness distributions to the images of strong maser spots and the
background sources using the \AIPS\ task {\it SAD} or {\it JMFIT}.

\section{ASTROMETRIC PROCEDURES}
\label{sec:procedure}

%--- How to obtain the parallax and absolute proper motion of masers

Data used for parallax and proper motion fits were residual position
differences between maser spots and background sources in eastward (\dx\
= $\Delta\alpha\cos\delta$) and northward (\dy\ = $\Delta\delta$)
directions.  These residual position differences are relative to the
coordinates used to correlate the VLBA data and shifts applied in
calibration.   The data were modeled by the parallax sinusoid in both
coordinates (determined by a single parameter, the parallax) and a
linear proper motion in each coordinate.  Because systematic errors
(owing to small uncompensated atmospheric delays and, in some cases,
varying maser and calibrator source structures) typically dominate over
thermal noise when measuring relative source positions, we added ``error
floors'' in quadrature to the formal position uncertainties. We used
different error floors for the \dx\ and \dy\ data and adjusted them to
yield post-fit residuals with reduced $\chi^2$ near unity for both
coordinates.

The apparent motions of the maser spots can be complicated by a
combination of spectral and spatial blending and changes in intensity.
Thus, for parallax fitting, one needs to find stable, unblended spots
and preferably use many maser spots to average out these effects.  We
selected maser spots with peak signal to noise ratio greater than 15 in
our data analysis. We considered maser spots at different epochs as
being from the same feature if their position separation from adjacent
epoch position was less than $\Delta t~\times $ 5 \masy, where $\Delta
t$ is the time gap in year between the two adjacent epochs.  Masers can
be time-variable with spot lifetimes of months to years. For solid parallax
fits, we selected only maser spots persisting over at least 1 year of
data to avoid large correlations between parallax and proper motion.  We
first fitted a parallax and proper motion to each maser spot relative to
each background source separately.
Since one expects the same parallax for all maser spots, we did a
combined solution (fitting with a single parallax parameter for all
maser spots, but allowing for different proper motions for each maser
spot) using all maser spots and background sources. We used this method
to fit parallax for all \hho\ maser sources.

%--- For 6.7 GHz methanol masers ionospheric delay calibration
Unlike the VLBI parallax measurement of 22 GHz \hho\ masers, for 6.7 GHz
\meth\ masers, the ionospheric delay is the dominant error source, which
can limit parallax accuracy. Since the parallaxes determined with
different background quasars often display a ``gradient'' on the sky, we
adopted a method in which we generate position data relative to an
``artificial quasar'' at the target maser position at each epoch.
Fitting parallax to these data can significantly improve parallax
accuracy.  Details of this method are described
in~\citet{2017AJ....154...63R}.  We have since improved the method by
fitting all the data in a single step (rather than first generating
artificial QSO data), and the results presented here model the
positional data of a maser spot relative to multiple quasars at epoch
$t$ as the sum of the maser's parallax and proper motion and a planar
``tilt,'' owing to ionospheric wedges, of the quasar positions about the
maser position: 

\begin{equation}
\Delta \theta _{s,q}^{x}(t) = \Pi ^{x} (t)+(\Delta \theta
_{s}^{x}-\Delta \theta _{q}^{x})+\mu _{s}^{x}\delta t +
S_{x}^{x}(t)\Theta _{q}^{x}+ S_{y}^{x}(t)\Theta _{q}^{y}
\label{eq:1step}
\end{equation}

In the Eq.~\ref{eq:1step}, $\Pi ^{x} (t)$ is the $x$-component of the
parallax shift; $\Delta\theta_{s}^{x}$ and $\Delta\theta _{q}^{x}$ are
constant offsets of maser spot, $s$, and QSO, $q$, from the position
used in correlation; $\mu _{s}^{x}\delta t$ is the $x$-component of
maser spot proper motion; $S_{x}^{x}(t)\Theta _{q}^{y}$ is the
$x$-position shift owing to an ionospheric wedge ``slope'' in the
$x$-direction (in mas~deg$^{-1}$) times the $x$-component of the
separation between the maser and QSO $q$ (in deg); $S_{y}^{x}(t)\Theta
_{q}^{y}$ is the $x$-position shift owing to an ionospheric wedge
``slope'' in the $y$-direction (in mas~deg$^{-1}$) times the
$y$-component of the separation between the maser and QSO $q$ (in deg)
at epoch $t$.  The offset of one maser spot was set to zero and held
constant, since one cannot solve for all $\Delta\theta^x$ terms with
relative position information. There is an analogous equation for the
$y$-coordinate. 
We used a Markov chain Monte Carlo approach to generate marginalized
probability distribution functions for all parameters, varying all
parameters simultaneously.

%--- How to obtain the absolute proper motion of exciting star
%--- Questions: is there only one exciting star in a HMSFR?

\hho\ maser spots are not usually distributed uniformly around their
central exciting stars, and their kinematics can be complicated by a
combination of expansion and rotation in outflows with speed of
typically tens of \kms~\citep{1992ApJ...393..149G}; this can limit the
accuracy of estimates of the absolute proper motion of the exciting
star(s).  In our cases, the maser sources have few spots and a simple,
narrow maser spectrum, and the motions of their central stars were
determined by averaging motions of spots persisting at least three
epochs, and then assigning a proper motion uncertainty of 10 \kms\ at
the measured distance. Unlike \hho\ masers, \meth\ masers move slowly,
typically a few \kms~\citep{2010ApJ...716.1356M}, so we expect only
small problems in estimating the accuracy of the proper motions of the
underlying stars, but we conservatively add an uncertainty of 5 \kms\
in quadrature to the formal uncertainties.
%
%--- V LSR estimation (Added)
The \VLSR\ of each source is estimated using the median value of
the maser emission velocities, which is consistent with that derived
from the Gaussian fits of the Galactic Ring Survey
(\citealt{2006ApJS..163..145J}) $^{13}$CO spectra.

\section{Results and Discussions}

We identify masers associated with the inner Perseus arm based on their
coincidence in Galactic longitude ($\ell$) and \VLSR\ ($v$) with a large
lane of emission at low absolute velocities that marks the Perseus arm
in CO and HI ($\ell-v$) diagrams. As shown in Figure~\ref{fig:co_lv},
$\ell-v$ loci of all the sources listed in Table~\ref{tab:pipm} are 
consistent with the trace of the Perseus arm, except for \Gfozesi, whose
velocity is about 15 \kms\ larger than that of other nearby sources, but whose
parallax indicates a Perseus arm association.
Table~\ref{tab:pipm} lists parallaxes (distances), proper motions and
\VLSR\ of all new masers sources reported in this paper together with
two sources from \citet{2013ApJ...775...79Z} in the inner Perseus
arm.
As recently highlighted for {\it Gaia} results by
\citet{2015PASP..127..994B}, estimating distance by simply inverting 
parallax can result in bias.  For a Gaussian parallax, $\Pi$, distribution, 
the inverse distribution is asymmetric with a peak at $1/\Pi$ and a tail
toward larger distances.  Thus, the expectation (mean) distance is greater 
than that of the peak.  Therefore, in Table~\ref{tab:pipm}, we present
two estimates of distance: 1) $1/\Pi$ with asymmetric uncertainties 
obtained by adding and subtracting the $1\sigma$ parallax uncertainty, and
2) that following Bailer-Jones, with a prior of an exponentially decreasing 
space density with a length scale of $L=6$ kpc (which implies a mode of 
12 kpc for the prior).  The differences between distance estimates for
any given source are small, since most uncertainties are smaller than $\pm25$\%
of the parallax and distances are comparable to the mode of the prior in
the Bailer-Jones approach, and we adopt the simpler method 1) values.

Figure~\ref{fig:pos} shows the locations of HMSFRs determined by
trigonometric parallaxes. Table~\ref{tab:pecmot} lists the peculiar
motions of the HMSFRs, caculated
using the Galactic parameters ($R_0$ =
8.31 kpc, $\Theta_0$ = 241 \kms) and the Solar Motion
values (\U\ =  10.5 \kms, \V\ = 14.4 \kms, and \W\ =
8.8 \kms), and assuming the universal rotation curve with three parameters
$a_1$ = 241 \kms, $a_2$ = 0.90 and $a_3$ = 1.46
from~\citet{2014ApJ...783..130R}.

%--- Pitch angle
Combining the eight sources presented here with two sources reported in
\citet{2013ApJ...775...79Z}, there are now ten sources tracing
the inner Perseus arm.
As shown in Figure~\ref{fig:pos}, these sources are consistent with
following a spiral from Galactic azimuth $\approx$  50 \deg\ to 110\deg\
(corresponding to Galactic longitude $\ell \approx$ 70 \deg\ to 30\deg)
and extending nearly 5 kpc in length.  Using a Bayesian fitting approach
that takes into account uncertainties in distance that map into both
$R$ and $\beta$~\citep{2014ApJ...783..130R} and is insensitive to
outliers \citep[see ``conservative formulation'']{2006OUP.book......S},
we estimate a pitch angle of 5\deg~ $\pm$ 4\deg\ for the inner portion
of the Perseus arm.  This is consistent with our previous estimates of
9\deg~$\pm$ 2\deg and 9\deg~ $\pm$ 1\deg\ determined from the sources
confined to the outer and all portions of the Perseus
arm~\citep{2013ApJ...775...79Z,2014ApJ...783..130R}, respectively.

%--- A Star Formation Gap in the Perseus arm
In our first paper on the inner Perseus arm~\citep{2013ApJ...775...79Z},
we suggested the existence of a gap in the distribution of high mass star-forming
regions in the Perseus arm based on the lack of masers and the paucity of massive 
young stellar objects in the Red MSX Source survey~\citep{2014MNRAS.437.1791U}.  While both suggest a low level
of \textit{active} star formation, the overall small number of 1.1-mm 
wavelength continuum sources detected in the Bolocam Galactic Plane
Survey shows that this part of the Perseus arm is also lacking dense
molecular cloud cores, the raw material of present and future star
formation \citep{2013ApJS..208...14G}.  It is clear from Figure 16 of
their study~\citep[see also Figure 6 of][] {2013ApJS..209....2S} and in the Bolocam source
catalog~\footnote{\url{http://irsa.ipac.caltech.edu/data/BOLOCAM_GPS/tables/bgps_v2.0.tbl}},
these authors only found a small number of compact mm-wavelength sources in longitudes
between 50\deg\ and 80\deg; i.e., a few tens as opposed to many
hundreds in longitude ranges of comparable width elsewhere in the plane.
In particular, many dense cores are found in the 30\deg\ $< \ell <$
50\deg\ range, in which the masers studied by us are located.
We also note that \citet{2017PASP..129i4102K} have confirmed the gap, based on
the clumpy distribution of HII regions detected in infrared emission with the
Wide-Field Infrared Survey Explorer (WISE) catalogued by~\citet{2014ApJS..212....1A}.
Furthermore, the recently updated HII region catalogue of~\citet{2018ApJS..234...33A}
as shown in Figure~\ref{fig:HII} clearly shows a much lower density of HII regions 
in 55\deg\ $< \ell <$ 100\deg\ than in the 30\deg\ $< \ell <$ 55\deg\ range.
The absence of massive star forming material in the Perseus gap supports
theories of spiral arm formation in segments
\citep{1966ApJ...146..810J,2012MNRAS.426..167G,2013ApJ...766...34D,2015MNRAS.454.2954B},
rather than global formation on a galaxy-wide scale as assumed by the
classic static density wave theory~\citep{1964ApJ...140..646L}. 
However, recently the possibility that reflection of spiral density waves
might also produce a segmented view of spiral arm star-formation has
been forwarded~\citep{2016ARA&A..54..667S}.

%
%--- Peculiar motions
Additional evidence for segmental arm formation can be found in maser
kinematics.  Table~\ref{tab:avg_pecmot} lists variance-weighted averages
of the peculiar motion components of three segments of the Perseus arm
with data from \citet{2014ApJ...783..130R} and this paper.
There is a significant difference in \Usbar\ between the inner and outer
segments (especially the segment in $\ell$ between 90\deg\ and 140\deg).
This further supports our finding that the Perseus arm is not a single
coherent feature, since the inner and the outer segments are moving apart and
may be independent.

\section{Summary}

We report trigonometric parallax and proper motion measurements of 6.7
GHz \meth\ masers and 22 GHz \hho\ masers from eight HMSFRs using the VLBA
observations as part of the BeSSeL Survey. The distances of these HMSFRs
combined with their Galactic coordinates, radial velocities, and proper
motions, allow us to assign them to the inner Perseus arm. These HMSFRs
are clustered between Galactic longitudes from $\approx$ 30\deg\ to
50\deg, a range for which other tracers indicated an enhanced level of
star-formation activity relative to other sections of the Perseus arm.

%+++ References
\clearpage

%+++ References
\clearpage

% \bibliographystyle{aasjournal}
% \bibliography{ref}

\begin{acknowledgements}

This work was partly supported by the National Science Foundation of
China under grant 11673051, the 100 Talents Project of the Chinese Academy
of Sciences, and the Key Laboratory for Radio Astronomy, Chinese Academy
of Sciences.

\end{acknowledgements}

\facilities{VLBA}

\software{DiFX~\citep{2007PASP..119..318D},
ParselTongue~\citep{2006ASPC..351..497K},
AIPS~\citep{1996ASPC..101...37V}}

\clearpage
%+++ TABLES

\begin{deluxetable}{lccllrl}
\tablecolumns{6}
\tablewidth{0pc}
\tabletypesize{\footnotesize} %8pt
\tablecaption{Parallaxes and Proper Motions of Maser Sources in the
Inner Perseus arm}
\tablehead{
\colhead{Source}  & \colhead{Parallax} & \colhead{Distance} & \colhead{\mux}    & \colhead{\muy}    & \colhead{\VLSR} &  \colhead{Reference}  \\
\colhead{name}    & \colhead{(mas)}    & \colhead{(kpc)}    & \colhead{(\masy)} & \colhead{(\masy)} & \colhead{(\kms)} &
}
  \colnumbers
\startdata
\Gthontw~(H)   & 0.076 $\pm$ 0.014  & $13.2_{-2.0}^{+3.0}$ ( $ 13.1^{+3.1}_{-2.1})$ & $-$2.80 $\pm$ 0.19  & $-$5.54 $\pm$ 0.19  &    24 $\pm$ 10      {(\phn21)}& This paper \\
\Gthtwse~(H)   & 0.103 $\pm$ 0.031  & $~9.7_{-2.2}^{+4.1}$ ( $ 10.0^{+5.1}_{-2.7})$ & $-$2.94 $\pm$ 0.24  & $-$6.07 $\pm$ 0.25  & \phn8 $\pm$ 15      {(\phn14)}& This paper \\
\Gfozefo~(C)   & 0.078 $\pm$ 0.013  & $12.8_{-1.8}^{+2.6}$ ( $ 12.8^{+2.7}_{-1.9})$ & $-$2.95 $\pm$ 0.09  & $-$5.48 $\pm$ 0.10  &    10 $\pm$ \phn5  {(\phn\phn\phn)}& This paper \\
\Gfozesi~(C)   & 0.080 $\pm$ 0.021  & $12.5_{-2.6}^{+4.4}$ ( $ 12.4^{+4.7}_{-2.8})$ & $-$2.69 $\pm$ 0.09  & $-$5.60 $\pm$ 0.25  &    31 $\pm$ \phn5  {(\phn33)}& This paper \\
\Gfotwze~(C)   & 0.071 $\pm$ 0.012  & $14.1_{-2.0}^{+2.9}$ ( $ 14.0^{+2.9}_{-2.1})$ & $-$2.40 $\pm$ 0.09  & $-$5.64 $\pm$ 0.11  &    12 $\pm$ \phn5  {(\phn\phn\phn)}& This paper \\
\Gfothon~(H) \tablenotemark{a} & 0.090 $\pm$ 0.006  & $11.1_{-0.7}^{+0.8}$ ( $ 11.1^{+0.8}_{-0.7})$ & $-$2.48 $\pm$ 0.15  & $-$5.27 $\pm$ 0.13  &    11 $\pm$ \phn5  {(\phn16)}& \citet{2013ApJ...775...79Z} \\
\Gfoeisi~(H)   & 0.093 $\pm$ 0.005  & $10.7_{-0.5}^{+0.6}$ ( $ 10.8^{+0.6}_{-0.6})$ & $-$2.89 $\pm$ 0.13  & $-$5.50 $\pm$ 0.13  &    18 $\pm$ \phn5  {(\phn18)}& \citet{2013ApJ...775...79Z} \\
\Gfonitw~(C)   & 0.113 $\pm$ 0.016  & $~8.8_{-1.1}^{+1.4}$ ( $ ~8.9^{+1.6}_{-1.2})$ & $-$2.73 $\pm$ 0.15  & $-$5.85 $\pm$ 0.19  & \phn0 $\pm$ \phn5  {(\phn\phn3)}& This paper \\
\Gfonifo~(C)   & 0.132 $\pm$ 0.031  & $~7.6_{-1.4}^{+2.3}$ ( $ ~7.9^{+3.1}_{-1.8})$ & $-$3.15 $\pm$ 0.17  & $-$4.49 $\pm$ 0.66  & $-$12 $\pm$ \phn5  {(--21)}& This paper \\
{ \Gsezeon~(C)}   & 0.136 $\pm$ 0.014  & $~7.3_{-0.7}^{+0.8}$ ( $ ~7.4^{+0.9}_{-0.7})$ & $-$2.88 $\pm$ 0.15  & $-$5.18 $\pm$ 0.18  & $-$23 $\pm$ \phn5  {(\phn\phn\phn)}& This paper \\
\enddata
\tablenotetext{a}{W49 N.}
\tablecomments{
Column 1 lists source names, where H and C in parentheses denote
\hho\ and \meth\ maser, respectively.
Column 3 lists two estimates of distance: the distance obtained by
inverting the parallax and, in parentheses, using the method of
\citet{2015PASP..127..994B}.
Column 4 to 5 give absolute proper motions in the eastward and northward directions.  Column 6 gives estimates of the \VLSR\ of the central star; 
the first values are from \S~\ref{sec:procedure} and the values in parentheses are from the nearest BGPS source 
\citep{2013ApJS..209....2S}, where empty parentheses indicate no source within 150\arcsec\ of the maser position.
\label{tab:pipm}
}
\end{deluxetable}

%--- Table for Peculiar Motions
\clearpage
\begin{deluxetable}{lrrr}
\tablecolumns{6}
\tablewidth{0pc}
\tabletypesize{\footnotesize} %8pt
\tablecaption{Peculiar Motions of Maser Sources in the Inner Perseus Arm}
\tablehead{
\colhead{Source}   & \colhead{\Us}    & \colhead{\Vs}    & \colhead{\Ws} \\
\colhead{name}     & \colhead{(\kms)} & \colhead{(\kms)} & \colhead{(\kms)}
}
\colnumbers
\startdata
\Gthontw           & --19.5 $\pm$   10.6  & --12.4 $\pm$ 70.8 &   6.7  $\pm$   12.7  \\
\Gthtwse           &   15.7 $\pm$   12.7  & --66.2 $\pm$ 98.8 &   1.9  $\pm$   13.4  \\
\Gfozefo           &  --5.2 $\pm$ \phn5.4 &   12.2 $\pm$ 60.2 &  15.6  $\pm$ \phn6.0  \\
\Gfozesi           & --21.5 $\pm$   10.0  &   19.2 $\pm$ 94.3 & --2.0  $\pm$ \phn9.8  \\
\Gfotwze           & --11.3 $\pm$ \phn5.7 &   44.8 $\pm$ 67.2 & --22.1  $\pm$ \phn8.9  \\
\Gfothon~(W49~N)   & --20.6 $\pm$ \phn6.3 & --25.1 $\pm$ 14.3 &  --3.4  $\pm$ \phn7.8  \\
\Gfoeisi           &  --7.7 $\pm$ \phn5.9 &    3.3 $\pm$ \phn9.3 &   7.3  $\pm$ \phn6.3  \\
\Gfonitw           &   11.6 $\pm$ \phn7.2 & --27.7 $\pm$ 26.1 &  --5.4  $\pm$ \phn7.2  \\
\Gfonifo           & --19.6 $\pm$   20.1  & --67.8 $\pm$ 34.0 &  32.8  $\pm$     15.7  \\
\Gsezeon           &   10.8 $\pm$ \phn5.9 &    0.5 $\pm$ 14.0 &  --2.2  $\pm$ \phn5.8  \\
\enddata
\tablecomments{Columns 1 lists source names,
columns 2 to 4 list peculiar motion components, where \Us, \Vs, \Ws\ are
directed toward the Galactic Center, in the direction of Galactic
rotation and toward the North Galactic Pole (NGP), respectively. The
peculiar motions were estimated using the Galactic parameters ($R_0$ =
8.31 kpc, $\Theta_0$ = 241 \kms) and the Solar Motion
values (\U\ =  10.5 \kms, \V\ = 14.4 \kms, and \W\ =
8.8 \kms), and assuming the universal rotation curve with three parameters
$a_1$ = 241 \kms, $a_2$ = 0.90 and $a_3$ = 1.46
from~\citet{2014ApJ...783..130R}.
\label{tab:pecmot}
}
\end{deluxetable}

%--- Table for Peculiar Motions
\clearpage
\begin{deluxetable}{crrrr}
\tablecolumns{6}
\tablewidth{0pc}
\tabletypesize{\footnotesize} %8pt
\tablecaption{Averaged Peculiar Motions of Maser Sources in the Inner Perseus Arm}
\tablehead{
\colhead{Segment $\ell$} & \colhead{N} & \colhead{\Usbar}  & \colhead{\Vsbar}  & \colhead{\Wsbar} \\
        \colhead{\deg}   &             &\colhead{(\kms)}  & \colhead{(\kms)} & \colhead{(\kms)}
 }
 \startdata
  30 -- 50  & 9  &  -8.4  $\pm$ 2.5  & -9.7  $\pm$ 8.3 &  2.6  $\pm$ 2.8 \\
  90 -- 140 & 16 &  14.1  $\pm$ 1.4  & -7.6  $\pm$ 1.4 & -1.7  $\pm$ 1.3 \\
 170 -- 245 & 9  &   3.5  $\pm$ 1.8  & -3.3  $\pm$ 2.0 & -1.2  $\pm$ 1.9 \\
\enddata
\tablecomments{Column 1 lists segments of the Perseus arm with 
different ranges of Galactic longitude as shown in
Figure~\ref{fig:pos}. Column 2 lists the number of values
averaged.
Columns 3 to 5 list variances weighted average peculiar motion components of
\Us, \Vs\ and \Ws\ for the arm segments using the data listed in
Table~\ref{tab:pecmot} and the data caculated from Table~1 in~\citet{2014ApJ...783..130R}.
\label{tab:avg_pecmot}
}
\end{deluxetable}

\clearpage

%--- Figure of parallax curves (1)
\begin{figure}[H]
 \includegraphics[angle=0,scale=0.44]{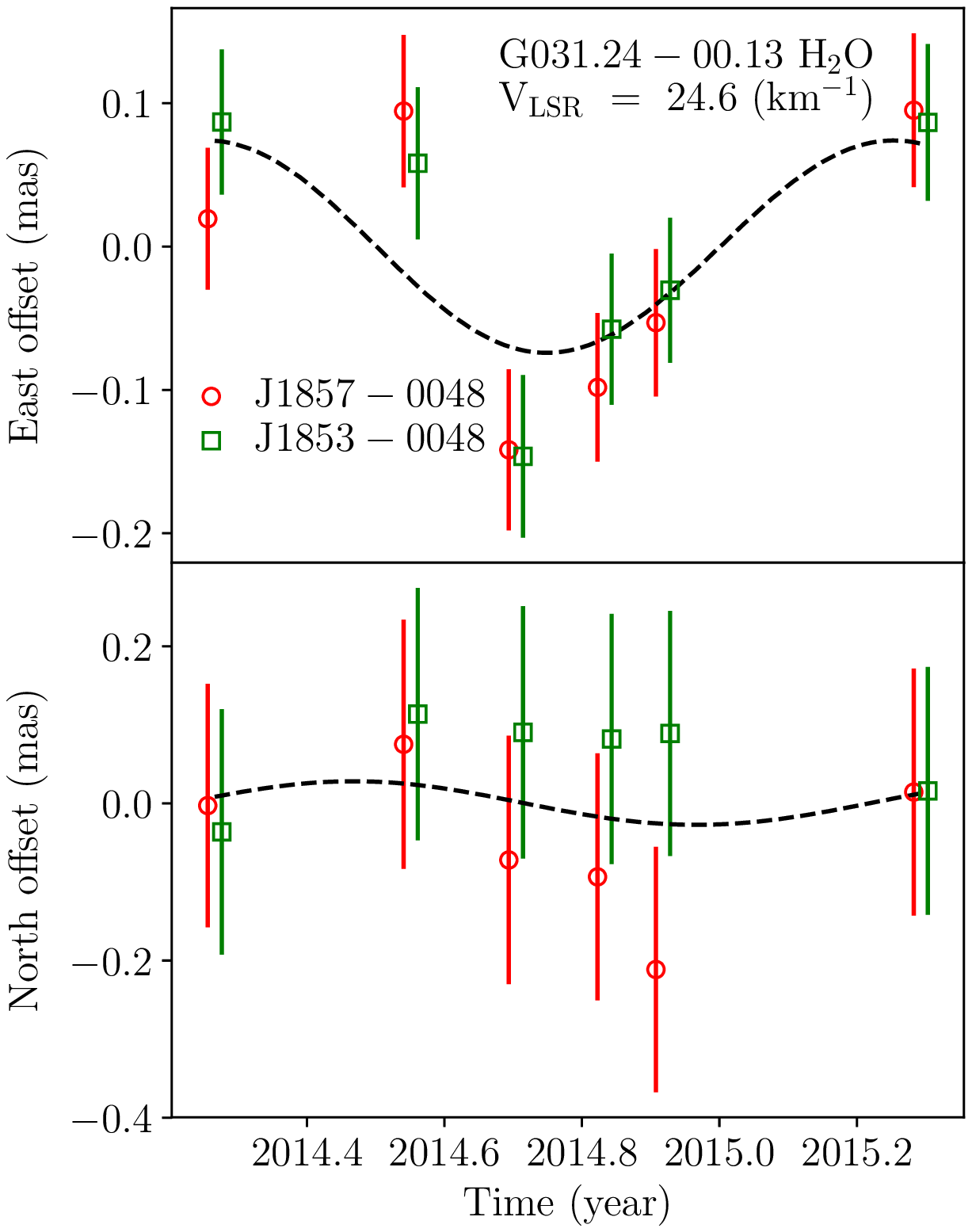}
 \includegraphics[angle=0,scale=0.44]{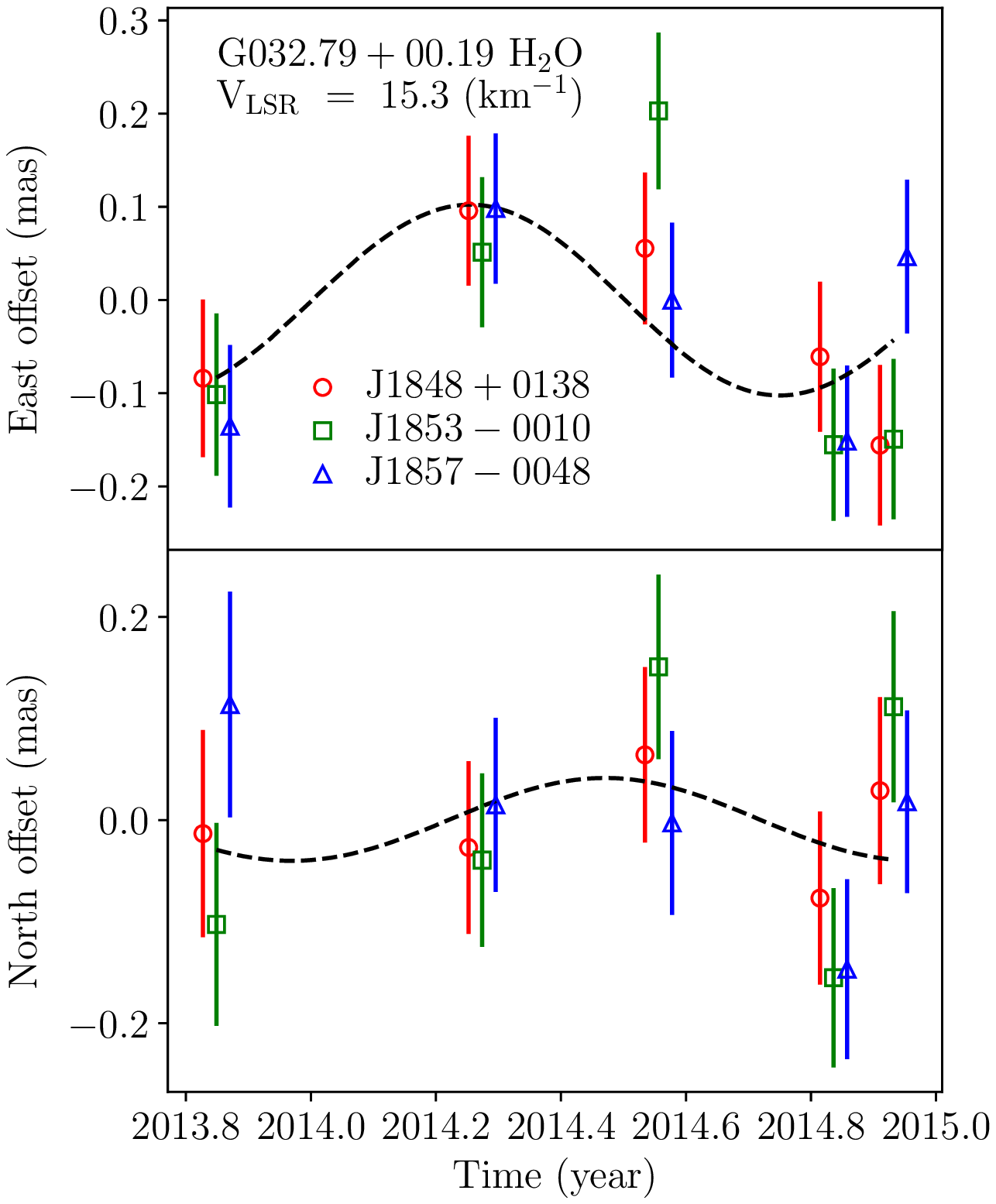}
 \includegraphics[angle=0,scale=0.44]{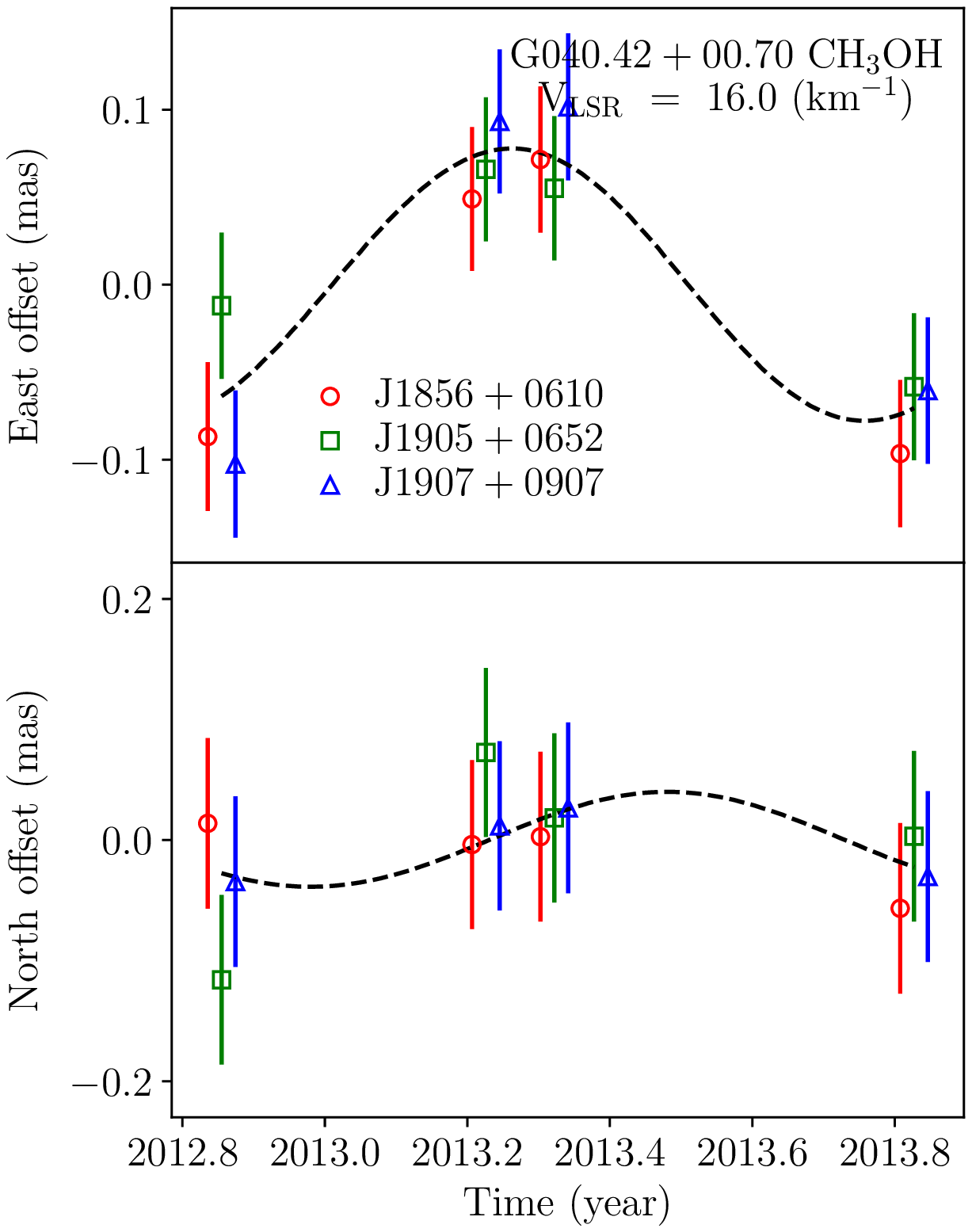}
 \includegraphics[angle=0,scale=0.44]{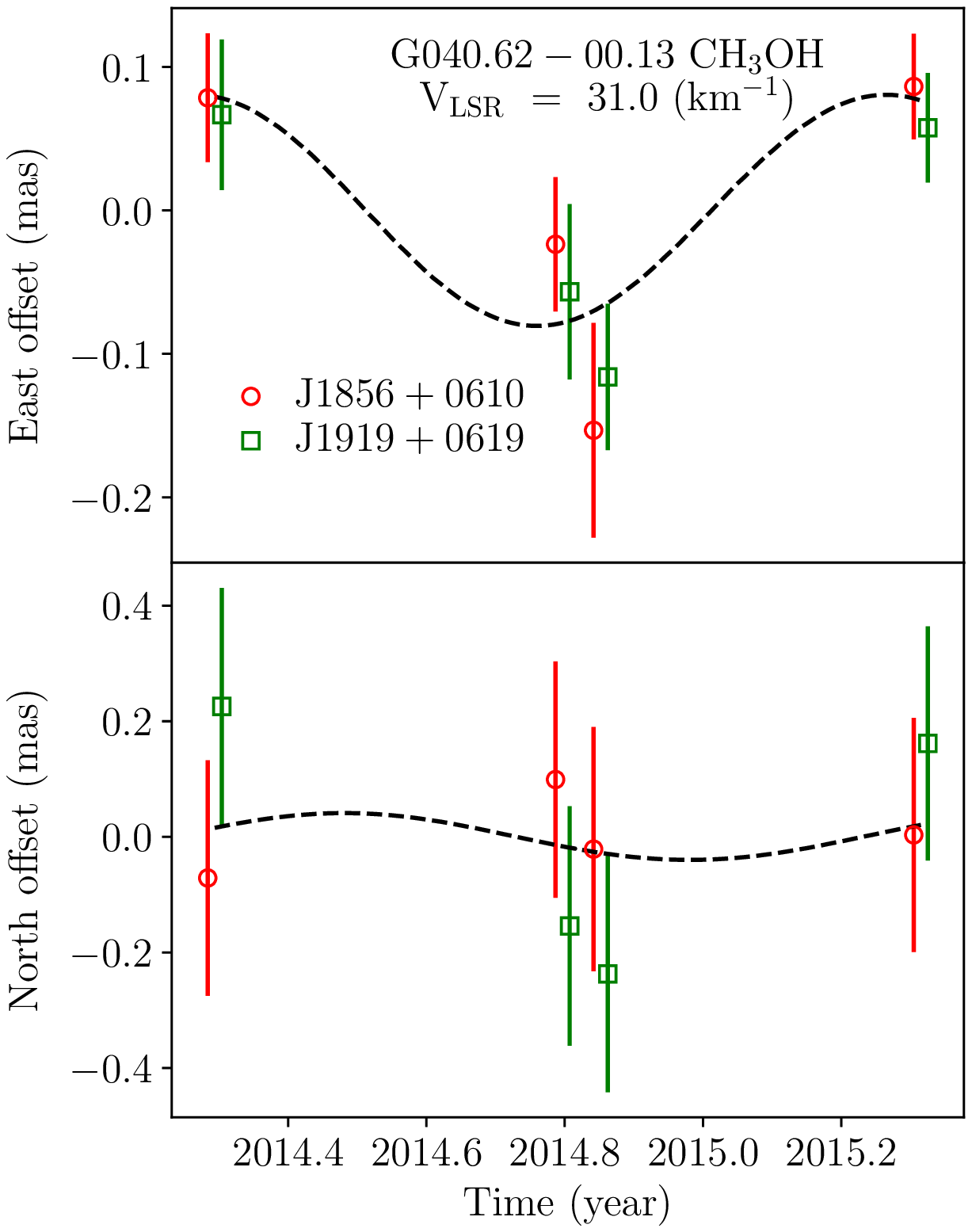}
 \includegraphics[angle=0,scale=0.44]{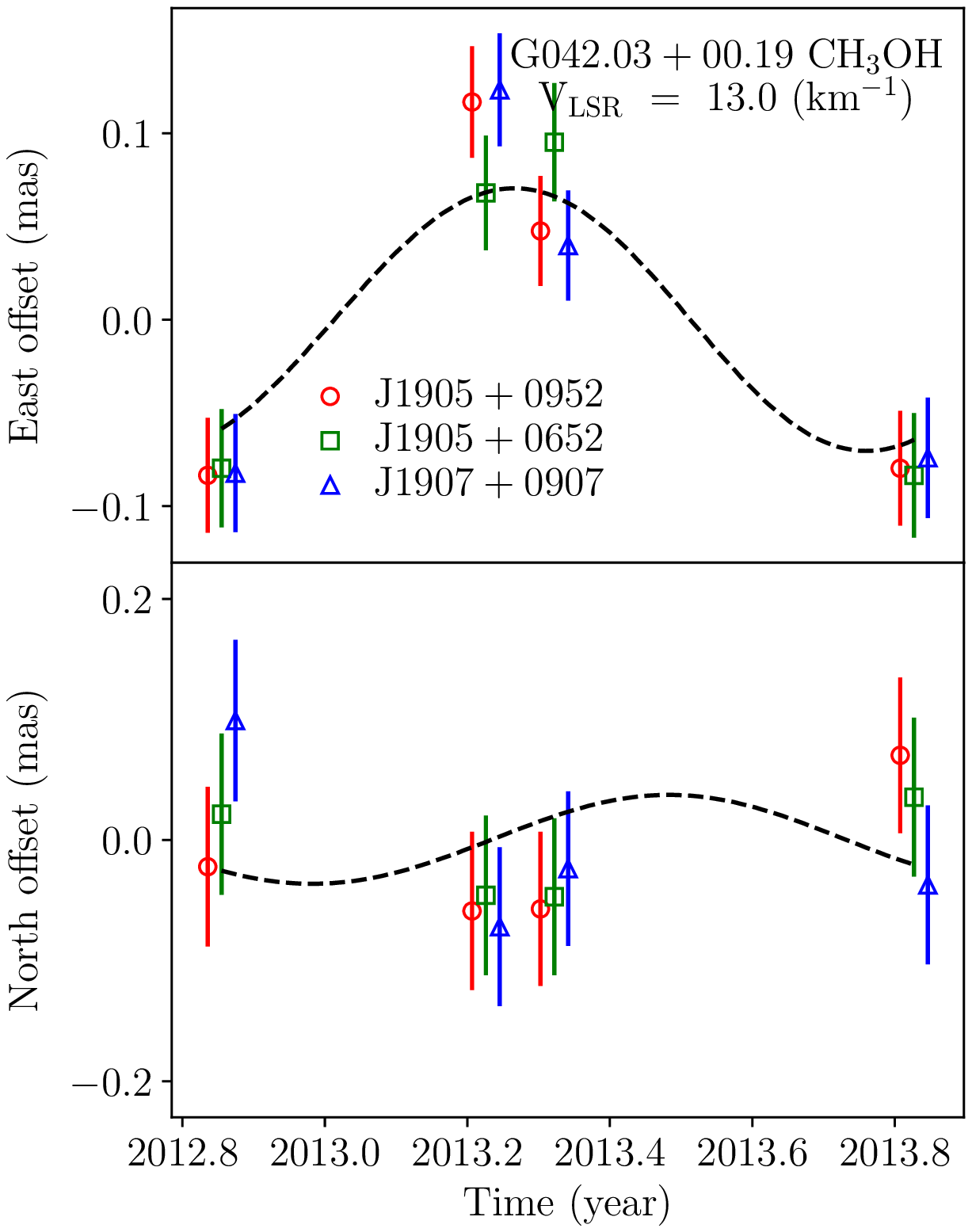}
 \includegraphics[angle=0,scale=0.44]{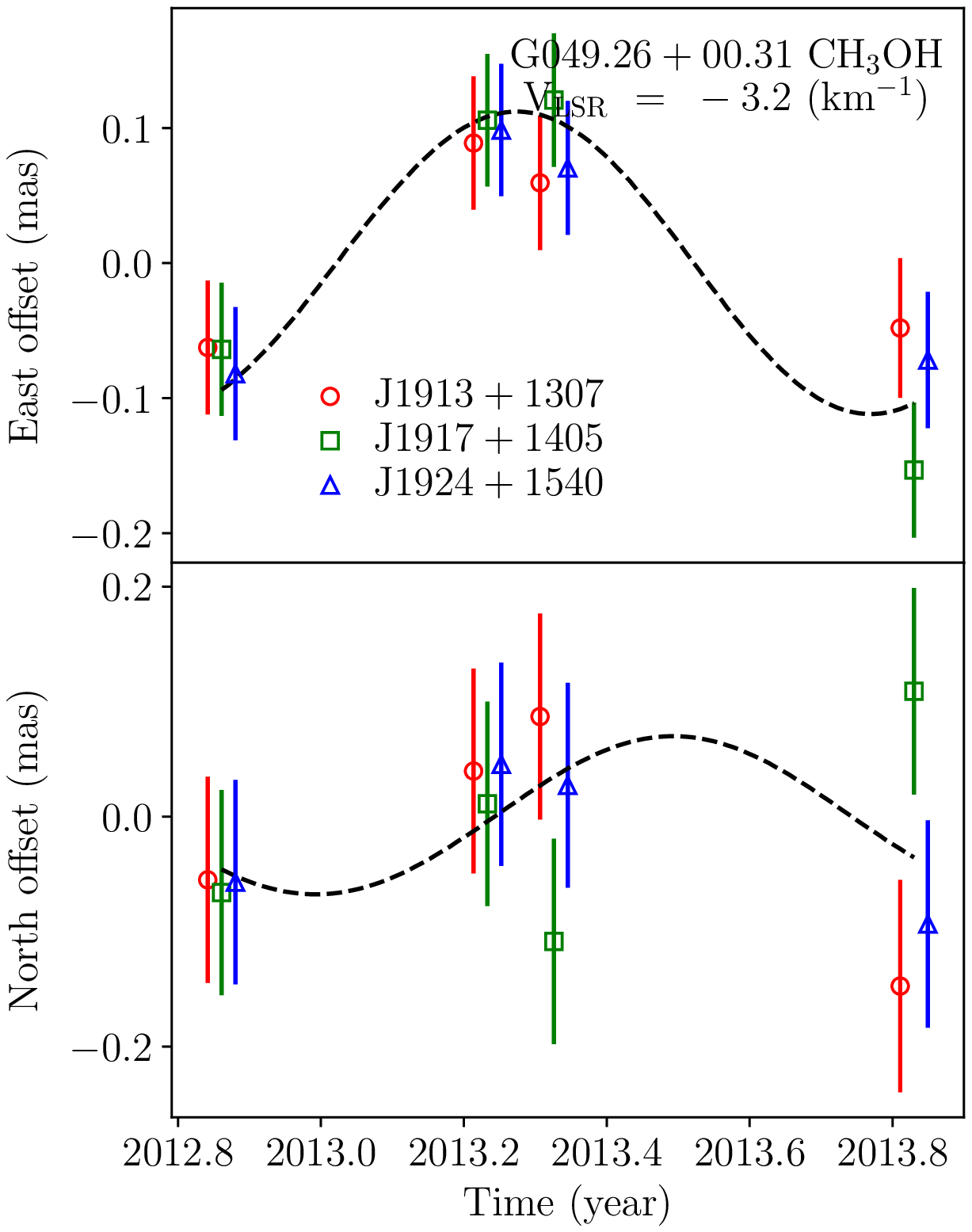}
 \includegraphics[angle=0,scale=0.44]{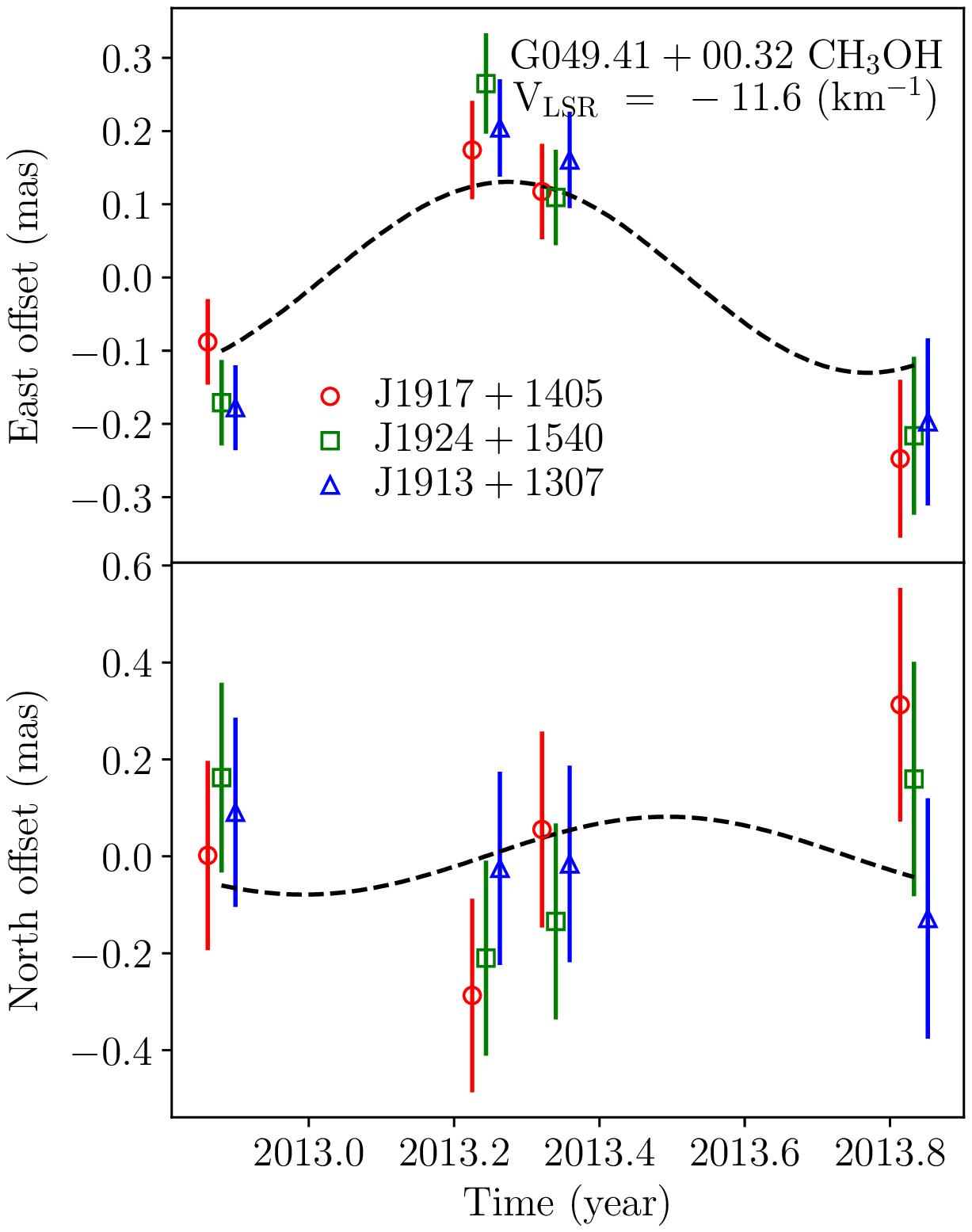}
 \hspace{0.1cm}
 \includegraphics[angle=0,scale=0.44]{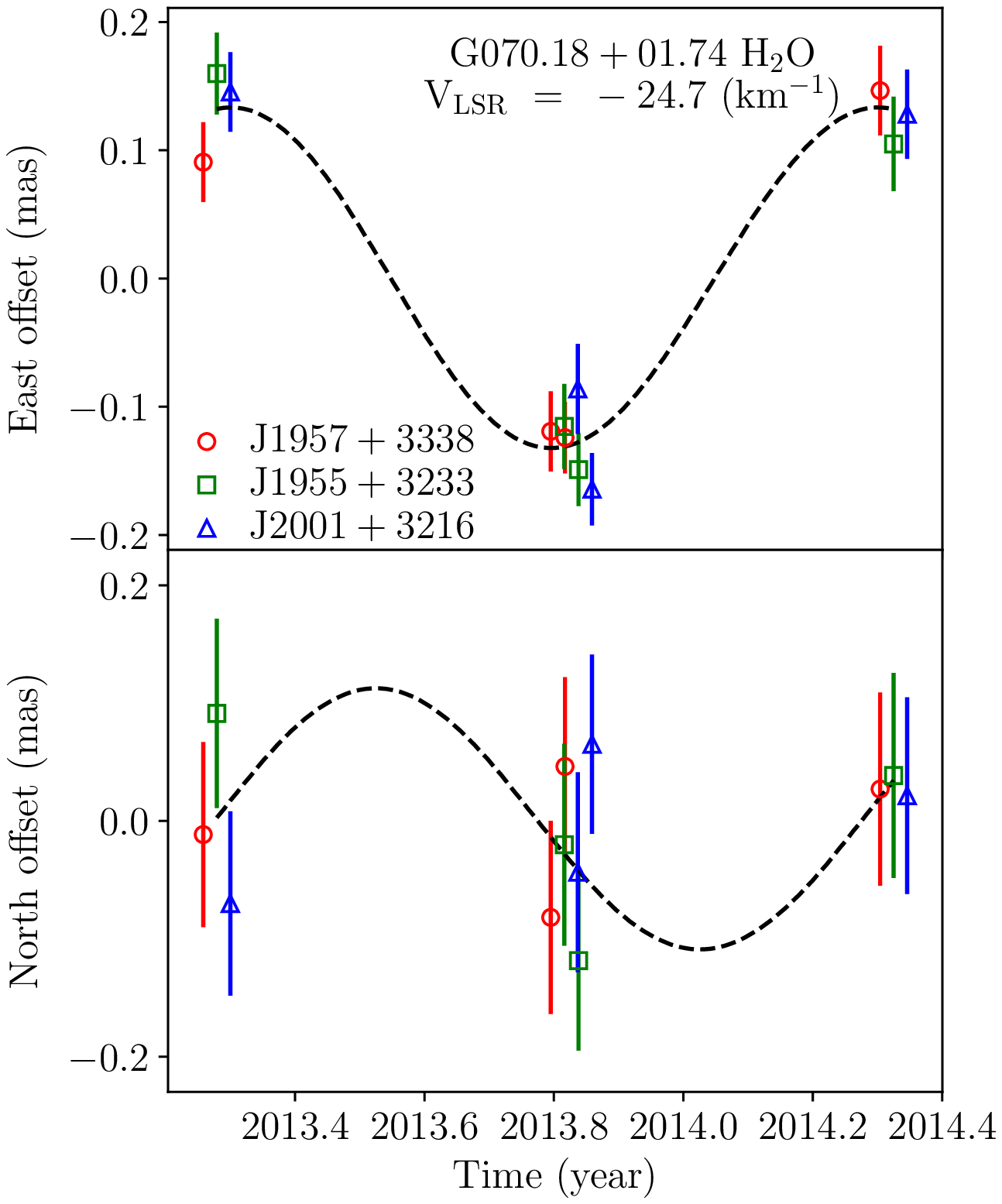}
 \caption{
 Proper motion-free Parallax signatures.  For \hho\ masers, the parallax is
 estimated by combining all the relative position data between the
 masers and QSOs.  For \meth\ masers, the parallax is estimated using the
 1-step method described in~\S~\ref{sec:procedure}.  In each panel, for
 clarity, the data of only one maser spot is plotted.  The QSO names used in
 the data fitting for each source, the \VLSR\ of the maser spot plotted, and
 the maser source names and its species are labeled.
 }
 \label{fig:para_curve1}.
\end{figure}

%--- Figure of HMSFRs distribution
\begin{figure}[H]
  \centering
  \includegraphics[angle=0,scale=0.90]{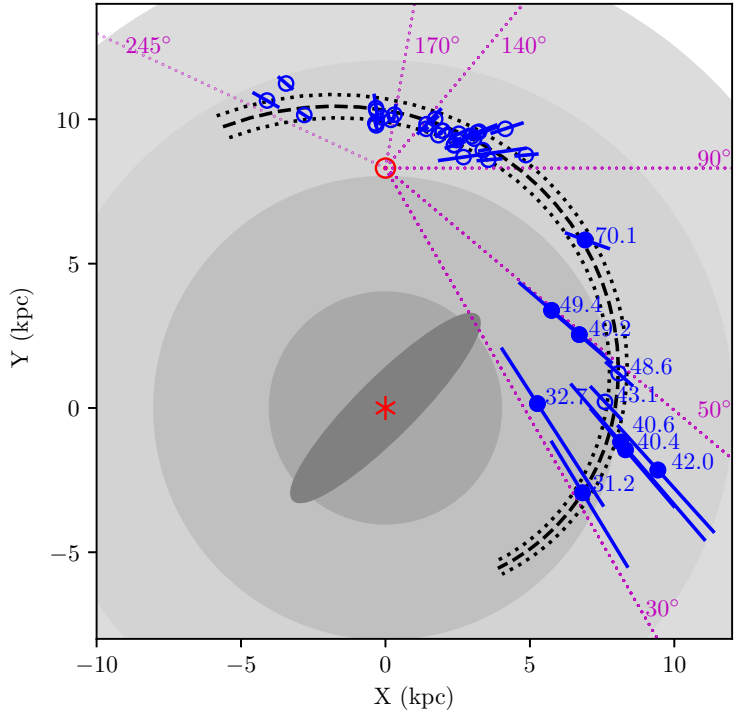}
  \caption{
  {\it Upper panel}: Locations ({\it markers with error bars}) of maser sources
  in the Perseus arm projected on the Galactic plane viewed from the North
  Galactic Pole. The values of locations are calculated using the trigonometric
  parallaxes. The filled circles are sources reported in this paper.  The
  Galactic Center ({\it asterisk}) is at (0, 0) and the Sun ({\it dot}) at (0,
  8.31) kpc.   The {\bf dashed} line is a section of the Perseus arm from a weighted fit
  to the points and corresponds to a global fitted pitch angle of 9\deg,
  and the dotted lines around the {\bf dashed} line denotes arm width 
  increasing with Galactocentric radius~\citep{2014ApJ...783..130R}.
  The Galactic longitude of each source is labeled near its
  location. The {\bf dotted straight} lines mark select Galactic longitudes discussed in
  the text.
  \label{fig:pos}}
\end{figure}

%--- Figure of CO l-v diagram
\begin{figure}[H]
  \centering
    \includegraphics[scale=1.25,angle=0]{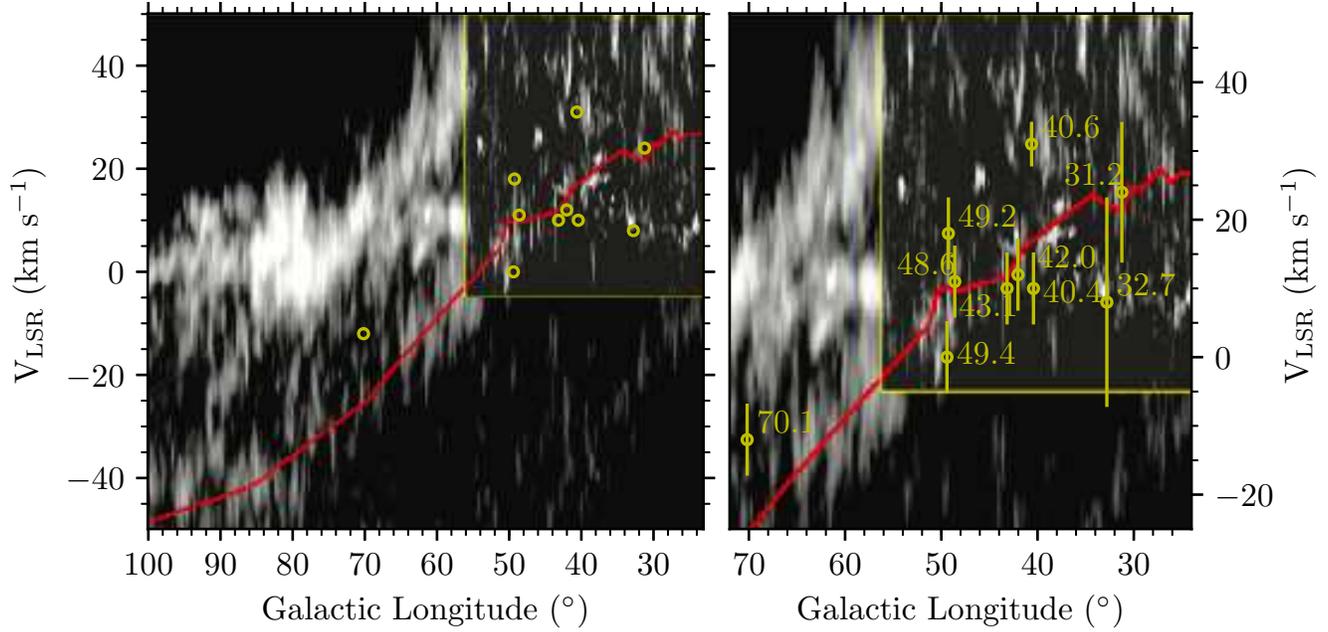}
    \caption{
    {\it Left panel}: Locations ({\it circles with error bar}) of the inner Perseus arm sources on
    a longitude-velocity diagram of CO emission
    survey~\citep{2001ApJ...547..792D}, which is adapted from Figure 8
    of~\citet{2016ApJ...823...77R}, where the Perseus arm in the first Galactic
    quadrant is traced by the {\it solid line}.  Within the yellow box at upper
    right, where the arm is most distant and within the solar circle,
    the~\citeauthor{2001ApJ...547..792D} survey is replaced by the higher
    angular resolution CO survey of~\citet{2006ApJS..163..145J}.
    {\it Right panel}: Zoom in plot from the {\it left panel} with Galactic 
    longitude of each source is labeled and \VLSR\ error bars shown.
    \label{fig:co_lv} }
\end{figure}

%--- Figure of HII regions
\begin{figure}[H]
  \centering
    \includegraphics[scale=0.95,angle=0]{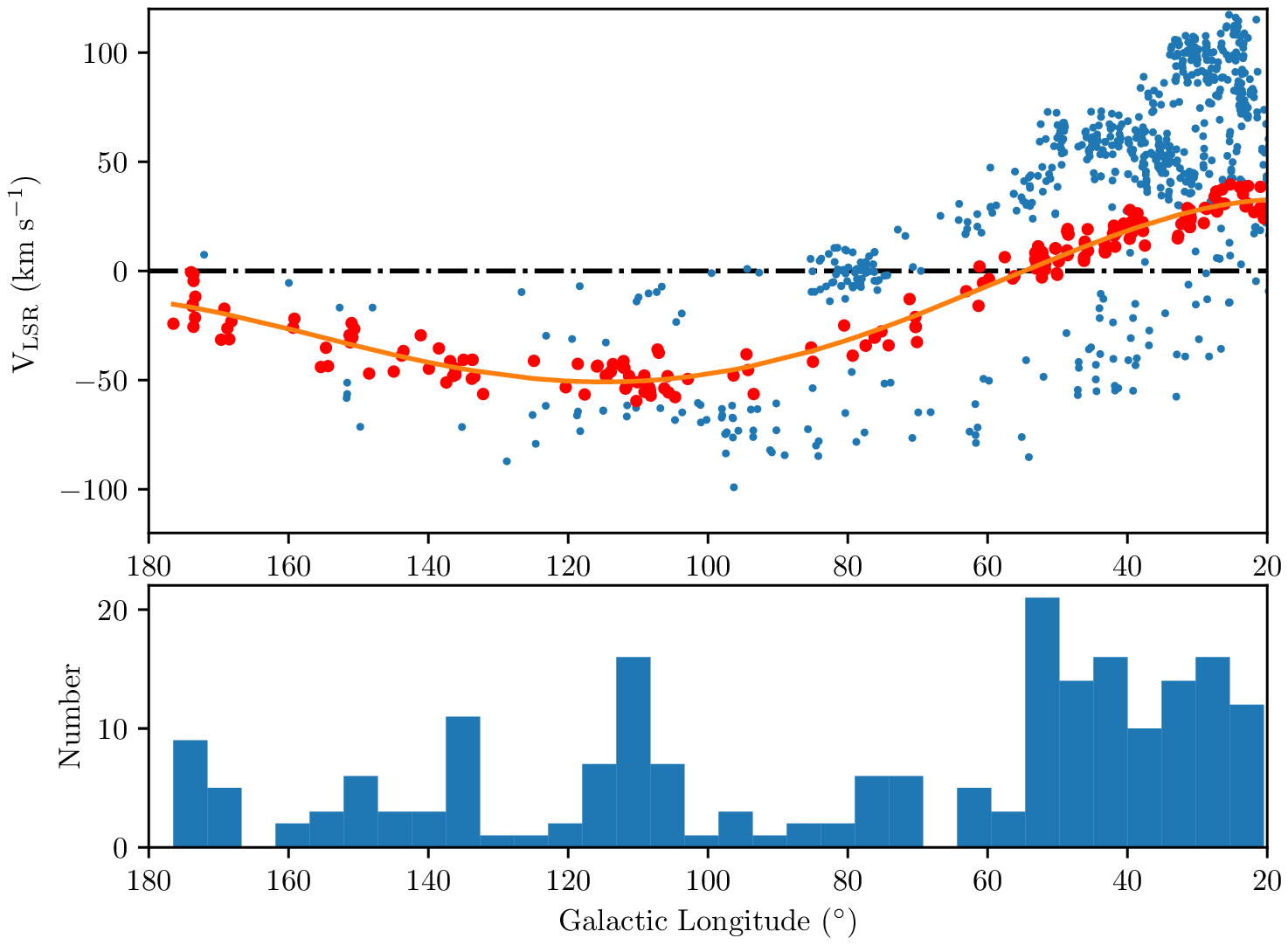}
    \caption{
    {\it Upper panel}: Distribution of WISE HII regions along the
    Perseus arm in the range of $\ell$ of 20--180\deg, where the HII
    regions are from the catalog on 
    \url{http://astro.phys.wvu.edu/wise/}(see also
    \citealt{2018ApJS..234...33A}).  The red dots denote the HII
    regions which can be assigned to the Perseus arm, which is traced by the solid
    curve of $\ell-v$ diagram as shown in Figure 8 of~\citet{2016ApJ...823...77R}; blue dots are sources in other spiral arms.
    {\it Lower panel}: Numbers of WISE HII regions on the Perseus arm versus Galactic
    longitude.
    \label{fig:HII} }
\end{figure}

\clearpage

\appendix
\setcounter{table}{0}
\renewcommand{\thetable}{A\arabic{table}}

%--- Table for VLBA observations
\begin{deluxetable}{lccl}
  \tablecolumns{5}
  \tablewidth{0pc}
  \tablecaption{VLBA Observations \label{tab:obs}}
  \tablehead{
   \colhead{Source} & \colhead{Maser}    & \colhead{Program} &\colhead{Epochs} \\
    \colhead{Name}  & \colhead{Species}  &  \colhead{Code}   &\colhead{(20yymmdd)}
  }
\colnumbers
  \startdata
 \Gthontw & \hho  &  BR198L &131104, 140407, 140720, 140914, 141031, 141212, 150417 \\
 \Gthtwse & \hho  &  BR198O &131106, 140410, 140722, 140920, 141101, 141206 \\
 \Gfozefo & \meth &  BR149O &121108, 130324, 130428, 131029 \\
 \Gfozesi & \meth &  BR149Q &121117, 130330, 130504, 131030 \\
 \Gfotwze & \meth &  BR149O &121108, 130324, 130428, 131029 \\
 \Gfonitw & \meth &  BR149P &121110, 130326, 130429, 131030 \\
 \Gfonifo & \meth &  BR149Q &121117, 130330, 130504, 131031 \\
 \Gsezeon & \meth &  BR149R &121118, 131026, 131102, 140429\\
\enddata
\end{deluxetable}

%--- Table for Source info
\startlongtable
\begin{deluxetable}{llcccccl}
  \tablecolumns{8}
  \tablewidth{0pc}
  \tablecaption{Positions and Brightnesses of the Target and Background Sources}
  \tablehead{
  \colhead{Source} & \colhead{R.A. (J2000)} & \colhead{Dec.  (J2000)}               & \colhead{$\theta_{sep}$} & \colhead{P.A.}    & \colhead{$S_p$}      & \colhead{\VLSR}  & \colhead{Beam} \\
  \colhead{      } & \colhead{(h~~~m~~~s)}  & \colhead{(\degr~~~\arcmin~~~\arcsec)} & \colhead{(\degr)}        & \colhead{(\degr)} & \colhead{(Jy/beam)}  & \colhead{(\kms)} & \colhead{(mas~~mas~~\degr)}
 }
 \colnumbers
\startdata
%
% G031.24
%
\Gthontw~{(21/13)}      & 18 48 45.08390 & $-$01 33 13.0980 & ...  & ...   &  9.4  &$+$26.0 & 1.4 $\times$ 0.6 @ \phn$-$8 \\
J1857$-$0048    & 18 57:51.35860 & $-$00:48:21.9496 & 2.4  &   72  & 0.011 &   ...  & 1.5 $\times$ 0.7 @ \phn$-$7 \\
J1853$-$0048    & 18 53:41.98920 & $-$00:48:54.3300 & 1.4  &   59  & 0.014 &   ...  & 1.4 $\times$ 0.6 @ \phn$-$10\\
J1846$-$0003    & 18 46:03.78500 & $-$00:03:38.2800 & 1.6  & --24  & 0.009 &   ...  & 1.4 $\times$ 0.7 @ \phn$-$8 \\
J1834$-$0301    & 18 34:14.07460 & $-$03:01:19.6270 & 3.9  &--112  & 0.045 &   ...  & 1.5 $\times$ 0.6 @ \phn$-$8 \\
&& && && & \\
%
% G032.79 (This source was analyzed by Choi, information should be added)
% --- reanalyzed by BZhang
\Gthtwse~{(8/8) }       & 18:50:30.7408  & $-$00:01:59.300  & ...  & ...    &  4.5   &$+$15.3 & 1.4 $\times$ 0.4 @ \phn$-$18 \\
J1853$-$0010    & 18:53:10.2692  & $-$00:10:50.740  & 0.7  & 102.5  & 0.005  &  ...   & 1.9 $\times$ 0.9 @ \phn$-$18 \\
J1857$-$0048    & 18:57:51.35860 & $-$00:48:21.9496 & 2.0  & 112.8  & 0.005  &  ...   & 1.9 $\times$ 0.8 @ \phn$-$13 \\
J1848$+$0138    & 18:48:21.81035 & $+$01:38:26.6322 & 1.8  & --17.8 & 0.014  &  ...   & 2.1 $\times$ 0.8 @ \phn$-$19  \\
&& && && & \\
%
% G040.42 (This is source was analyzed by Hu Bo, detailed info losted!),
% new detailed info was added by zb with re-anlayzing the 1st epoch data
%
%\Gfozefo   & 19:02:39.5843 & +06:59:14.674 &...&... && & \\ (HuBo)
\Gfozefo~{ (3/1)}   & 19:02:39.6194 & +06:59:09.052 &...  &...   & 4.6 & $+16.0$ & 5.1 $\times$ 1.7 @ \phn$-$21 \\
J1856+0610 & 18:56:31.8390 & +06:10:16.768 & 1.7 &--118 & 0.065 &   ... & 4.6 $\times$ 1.4 @ \phn$-$19 \\
J1854+0542 & 18:54:36.1284 & +05:42:59.307 & 2.4 &--122 & 0.037 &   ... & 4.7 $\times$ 1.7 @ \phn$-$22 \\
J1907+0907 & 19:07:41.9634 & +09:07:12.397 & 2.5 & 30   & 0.101 &   ... & 4.5 $\times$ 1.5 @ \phn$-$16 \\
J1905+0652 & 19:05:21.2105 & +06:52:10.780 & 0.7 & 100  & 0.073 &   ... & 4.8 $\times$ 1.4 @ \phn$-$19\\
&& && && & \\
%
% G040.62 (from Yuanwei)
\Gfozesi~{ (2/2)}   & 19:06:01.62870 & +06:46:36.1400 & ...  & ... &  4.7  & +31.0 &3.9 $\times$ 2.3 @ 13   \\
J1905+0652 & 19:05:21.21048 & +06:52:10.7803 & 0.2  &--61 & 0.092 &   ... &7.4 $\times$ 3.5 @ 52   \\
J1856+0610 & 18:56:31.83880 & +06:10:16.7650 & 2.4  &--104& 0.186 &   ... &7.7 $\times$ 3.5 @ 50   \\
J1912+0518 & 19:12:54.25770 & +05:18:00.4220 & 2.3  &--131& 0.081 &   ... &7.5 $\times$ 3.5 @ 51   \\
J1919+0619 & 19:19:17.35020 & +06:19:42.7700 & 3.3  & 98  & 0.035 &   ... &7.4 $\times$ 3.5 @ 51   \\
&& && && & \\
%
% G042.03 (This is source was analyzed by Hu Bo, detailed info losted!)
% new detailed info was added by zb with re-anlayzing the 1st epoch data
%
%\Gfotwze    & 19:07:28.1536 & +08:10:48.085 & ... & ...  &   & & \\
\Gfotwze~{ (3/3)}    & 19:07:28.1839 & +08:10:53.435 & ... & ...  & 3.1   & $+13.0$ & 5.1 $\times$ 1.6 @ \phn$-$20 \\
J1905+0952   & 19:05:39.8989 & +09:52:08.407 & 1.7 & --15 & 0.090 & ...     & 5.7 $\times$ 2.4 @ \phn$-$18 \\
J1913+0932   & 19:13:24.0254 & +09:32:45.379 & 2.0 &   47 & 0.019 & ...     & 5.6 $\times$ 2.3 @ \phn$-$19   \\
J1907+0907   & 19:07:41.9634 & +09:07:12.397 & 0.9 &    3 & 0.118 & ...     & 5.7 $\times$ 2.3 @ \phn$-$19 \\
J1905+0652   & 19:05:21.2105 & +06:52:10.780 & 1.4 &--158 & 0.075 & ...     & 5.7 $\times$ 2.4 @ \phn$-$18 \\
&& && && & \\
%
% G049.26
%
\Gfonitw~{ (4/4)}     & 19 20 44.8579 &  +14 38 26.555 & ... &  ... &   3.850  & --5.0 & 2.8 $\times$ 1.1 @ \phn$-$8 \\
  J1913+1307 & 19 13 14.0063 &  +13 07 47.343 & 2.4 &--130 &   0.069  &  ... & 4.1 $\times$ 1.5 @ \phn$-$19 \\
  J1917+1405 & 19 17 18.0637 &  +14 05 09.776 & 1.0 &--124 &   0.060  &  ... & 4.0 $\times$ 1.4 @ \phn$-$19 \\
  J1922+1530 & 19 22 34.6994 &  +15 30 10.028 & 1.0 &   27 &   0.061  &  ... & 3.5 $\times$ 1.5 @ \phn$-$20 \\
  J1924+1540 & 19 24 39.4558 &  +15 40 43.947 & 1.4 &   42 &   0.357  &  ... & 4.0 $\times$ 1.5 @ \phn$-$19 \\
&& && && & \\
%
% G049.41
%
\Gfonifo~{ (5/5)}& 19 20 59.1958 & +14 46 49.291 &...  & ...  &  1.160  & --11.9 &5.8 $\times$ 2.3  @ \phn$-$22 \\
  J1913+1307 & 19 13 14.0063 & +13 07 47.339 &2.5  &--131 &  0.073  &  ...   &7.7 $\times$ 3.3  @ \phn$-$22 \\
  J1917+1405 & 19 17 18.0638 & +14 05 09.774 &1.1  &--128 &  0.074  &  ...   &7.2 $\times$ 3.0  @ \phn$-$25 \\
  J1922+1530 & 19 22 34.6994 & +15 30 10.030 &0.8  &  28  &  0.099  &  ...   &7.2 $\times$ 3.0  @ \phn$-$25 \\
  J1924+1540 & 19 24 39.4558 & +15 40 43.944 &1.3  &  45  &  0.380  &  ...   &7.1 $\times$ 3.2  @ \phn$-$22 \\
&& && && & \\
{\Gsezeon~ (4/2)}& 20 00 54.1360 & +33 31 31.031 &...  & ...  & 3.100  & --11.9 &2.7 $\times$ 1.5  @ \phn$+$180 \\
 J2001+3323 & 20 01 42.20694 & +33 23 44.7461 &0.2 &  128 &  0.089  &  ...   &2.6 $\times$ 1.4  @ \phn$+$177\\
 J1957+3338 & 19 57 40.54974 & +33 38 27.9429 &0.7 &--80  &  0.133  &  ...   &2.7 $\times$ 1.5  @ \phn$+$180\\
 J1955+3233 & 19 55 56.35092 & +32 33 04.5060 &1.4 &--133 &  0.031  &  ...   &2.7 $\times$ 1.6  @ \phn$+$180\\
 J2001+3216 & 20 01 16.77409 & +32 16 46.9435 &1.3 &  176 &  0.032  &  ...   &2.5 $\times$ 1.5  @ \phn$+$4    
\enddata
\tablecomments{
The fourth and seventh columns give the peak brightnesses
($S_p$) and \VLSR\ of reference maser spot.  The fifth and sixth columns
give the separations ($\theta_{sep})$ and position angles (P.A.) east of
north of the background sources relative to the maser.  The last column gives the
FWHM size and P.A. of the Gaussian restoring beam.  All calibrators are
from \url{http://astrogeo.org}.%}
The two numbers in parentheses after maser source name denote numbers of maser spots with life time
longer than 1 yr and used in parallax estimation, respectively.
}
  \label{tab:src}
\end{deluxetable}

\end{document}